\documentclass[aps,prx,reprint,showpacs,nosuperscriptaddress,floatfix]{revtex4-2}
\usepackage[utf8]{inputenc}
\usepackage{graphicx}

\usepackage{hyperref}
\hypersetup{
colorlinks=true,final=true,
        linkcolor=blue,
        citecolor=blue,
        filecolor=blue,
        urlcolor=blue,
}

\begin{document}

\title{Electronic correlations, layer distinction, and electron doping in the alternating single-layer trilayer {L}a$_{3}${N}i$_{2}${O}$_{7}$ polymorph}
\author{Harrison LaBollita}
\email{hlabolli@asu.edu}
\affiliation{Department of Physics, Arizona State University, Tempe, AZ 85287, USA}
\author{Soumen Bag}
\affiliation{Department of Physics, Arizona State University, Tempe, AZ 85287, USA}
\author{Jesse Kapeghian}
\affiliation{Department of Physics, Arizona State University, Tempe, AZ 85287, USA}
\author{Antia S. Botana}
\affiliation{Department of Physics, Arizona State University, Tempe, AZ 85287, USA}

\begin{abstract}
We employ a density-functional theory plus dynamical mean-field theory framework to investigate the correlated electronic structure of the alternating single-layer trilayer (1313) polymorph of La$_3$Ni$_2$O$_7$, that becomes superconducting under pressure. At ambient pressure, the single-layer is in a Mott insulating regime and the low-energy physics is dominated by the trilayer block. Under pressure, the gap in the single-layer block closes due to orbital-selective physics, enabling charge transfer into the trilayer block. This change in effective doping of the trilayer block could be linked to the higher T$_c$ obtained in La$_3$Ni$_2$O$_7$-1313 ($\sim$ 80 K) when compared to the nominal trilayer La$_4$Ni$_3$O$_{10}$ compound ($\sim$ 30 K). We conclude that correlation-driven layer differentiation is crucial in the La$_3$Ni$_2$O$_7$-1313 polymorph and that its low-energy physics aligns closely with the trilayer La$_4$Ni$_3$O$_{10}$ compound (in spite of the apparent differences in nominal filling) rather than with the conventional bilayer La$_3$Ni$_2$O$_7$.
\end{abstract}

\maketitle

\section{\label{sec:intro}introduction}
The discovery of superconductivity in hole-doped infinite-layer nickelates $R$NiO$_2$ ($R$ = rare-earth) ~\cite{Li2019superconductivity,Osada2020superconducting, Osada2021nickelate, Zeng2021superconductivity} marked the culmination of a multi-decade pursuit for nickelate analogs to the  cuprates~\cite{Keimer2015From,Anisimov1999electronic, Pickett2004infinite}. The subsequent realization of superconductivity in a chemically undoped quintuple-layer nickelate Nd$_6$Ni$_5$O$_{12}$~\cite{Pan2021superconductivity} (with a similar T$_c$ $\sim$ 15 K) established the first family of nickelate superconductors represented by the general chemical formula $R_{n+1}$Ni$_n$O$_{2n+2}$ \cite{Lacorre1992, Polatvets2006la326, Poltavets2007crystal} ($n=2,\dots,\infty$). The materials in this family contain $n$-cuprate-like NiO$_2$ planes and an average Ni-3$d$ filling that can be tuned with the number of layers ($n$) corresponding to $d^{9-1/n}$ \cite{Labollita2021electronic}. 

The layered nickelates of the $R_{n+1}$Ni$_n$O$_{2n+2}$ family are derived via oxygen reduction from the parent $R_{n+1}$Ni$_n$O$_{3n+1}$ Ruddlesden-Popper (RP) compounds which host $n$-perovskite layers separated by rocksalt $R$-O spacer layers, with an average Ni-$3d$ filling of $d^{7+1/n}$~\cite{Greeenblatt1997ruddlesden, Ling2000neutron, Li2020epitaxial, Lei2017constructing, Pan2022synthesis}. Last year, superconductivity was realized under pressure in this parent RP family in bilayer La$_3$Ni$_2$O$_7$ with a record T$_c$ $\sim$ 80 K at 14 GPa \cite{sun2023superconductivity,zhang2023exps,hou2023emergence,li2024pressuredriven}, as well as in the trilayer La$_4$Ni$_3$O$_{10}$ material with a T$_c$ $\sim$ 30 K at $20-30$ GPa \cite{li2023signature, zhang2023superconductivity, zhu2024superconductivity}. These discoveries revealed a second, albeit related, family of nickelate superconductors that exhibits higher superconducting transition temperatures than their reduced counterparts, stimulating a plethora of experimental~\cite{Wang2024pressure,wang2023i4mmmexp,filamentary,li2023design} and theoretical~\cite{zhang2023electronic, gu2023effective, chen2023critical, lechermann2023electronic, christiansson2023correlated, luo2023bilayer, shen2023effective, zhang2023electronic, yang2023possible, liu2023swave, zhang2023structural, geisler2023structure, labollita2023electronic, labollita2024electronic, qu2023bilayer, yang2023minimal, zhang2023trends, lu2023superconductivity, tian2023correlation, ryee2024quenched, vortex2023huang, jiang2023screening, liao2023correlations, liao2023interlayer,oh2023tj,qin2023singlets, sakakibara2023hubbard, shilenko2023, wu2023zhangrice,leonov2024electronic, wang2024nonfermi, zhang2024prediction,tian2024effective,ouyang2024hund,craco2024strange} work.

Recently, a new polymorph of the bilayer RP phase La$_3$Ni$_2$O$_7$  was reported, characterized by a novel stacking sequence where alternating single-layer and trilayer blocks of NiO$_6$ octahedra form a `1313' configuration \cite{chen2023polymorphism, puphal2023unconventional,Wang2024longrange}, deviating from the typical uniform bilayer stacking of perovskite blocks (2222). Superconductivity in this new La$_3$Ni$_2$O$_7$-1313 polymorph has also been reported under pressure with a similar T$_c$ to that of the conventional bilayer nickelate La$_3$Ni$_2$O$_7$-2222~\cite{puphal2023unconventional}. Some theory work has been performed to analyze the electronic structure of this new polymorph. Density-functional theory plus dynamical mean-field theory (DFT+DMFT) has been employed to compare the electronic structure of the 2222 and 1313 polymorphs \cite{lechermann2024electronic}. Even though both phases display Ni-$d_{z^{2}}$ flat-band character at low-energy, the 1313 is distinct in that it exhibits significant layer selectivity, rendering especially the single-layer part to be insulating at ambient pressure. At high pressure, this layer selectivity weakens and the 1313 fermiology seems to display `arcs' reminiscent of the cuprates. In contrast, constrained RPA calculations~\cite{zhang2024electronic} conclude that the leading pairing instability in La$_3$Ni$_2$O$_7$-1313 is restricted to the single-layer block. DFT calculations comparing the optical conductivity of different members of the RP family indicate similarities between La$_3$Ni$_2$O$_7$-1313 and the trilayer RP  La$_4$Ni$_3$O$_{10}$ \cite{geisler2024opticalpropertieselectroniccorrelations}. Since La$_3$Ni$_2$O$_7$-1313 combines the single-layer La$_2$NiO$_4$~\cite{Lander1989,Rodriguez-Carvajal1991} and trilayer La$_4$Ni$_3$O$_{10}$~\cite{Zhang2020oxygen} structures, an important question to analyze systematically is whether such a structural superposition extends to the electronic structure as well.

In this work, we employ a DFT+DMFT framework to study the correlated electronic structure of the La$_3$Ni$_2$O$_7$-1313 polymorph under pressure and analyze its resemblance to the underlying single-layer and trilayer structural constituents. At ambient pressure, a Mott insulating state is obtained in the single-layer block with the low-energy physics being consequently dominated by the trilayer unit. Under pressure, the gap in the single-layer block closes and the trilayer block is effectively electron-doped. This effective doping may be related to the increased superconducting transition temperature of La$_3$Ni$_2$O$_7$-1313 relative to the trilayer RP counterpart La$_4$Ni$_3$O$_{10}$. Our findings suggest that a significant correlation-driven layer differentiation is present in La$_3$Ni$_2$O$_7$-1313 and we show that the low-energy correlated physics governing this polymorph aligns more closely with the (electron-doped) trilayer RP phase than with the conventional La$_3$Ni$_2$O$_7$-2222 nickelate.

\section{\label{sec:methods}Methodology}
The combination of density-functional theory (DFT) and dynamical mean-field theory (DMFT) was used to study the low-energy correlated electronic structure of La$_{3}$Ni$_{2}$O$_{7}$-1313 and its structural constituents: La$_{2}$NiO$_{4}$ and La$_{4}$Ni$_{3}$O$_{10}$.  Density-functional theory calculations were performed using the all-electron, full potential code \textsc{wien}2k~\cite{Blaha2020wien2k}. The correlated space was spanned by the Ni-$e_{g}$ \{$d_{x^{2}-y^{2}}$, $d_{z^{2}}$\} orbitals in our calculations, which we obtained by downfolding the Kohn-Sham DFT bands onto maximally-localized Wannier functions (MLWFs)~\cite{Arash2014wannier90, Kunes2010wien2wannier} corresponding to the Ni-$e_{g}$ states. Single-site dynamical mean-field theory calculations were performed using the \texttt{solid\_dmft}~\cite{soliddmft} package, built on the TRIQS software library~\cite{Parcollet2015triqs}. A continuous-time hybridization expansion (CT-HYB) solver as implemented in TRIQS/cthyb~\cite{Priyanka2016cthyb} was used to solve the impurity problem(s). Interactions were modeled by a Hubbard-Kanamori Hamiltonian including all spin-flip and pair hopping terms with $U = 5$ eV, $J_{\mathrm{H}}$ = 1 eV, and $U' = U - 2J$. Our choice of $U$ and $J_{\mathrm{H}}$ is consistent with previous theoretical works on the nickelates~\cite{ouyang2024hund,craco2024strange,tian2024effective}. Moreover, our results remain consistent within the range of $U = 4 - 6$ eV, with $J_{\mathrm{H}}$ fixed at 1 eV. All calculations were performed at an inverse electronic temperature corresponding to $\beta = 100$ eV$^{-1}$ and in the paramagnetic state. Instead of full charge self-consistency, our calculations adopt the simpler ``one-shot'' approach, which has demonstrated success in yielding valuable insights into the correlated physics of various transition-metal oxides~\cite{Mravlje2011coherence,Wang2012covalency, park2014total, Hampel2020effect} including nickelates~\cite{Karp2020manybody,Kitatani2020nickelate,labollita02022correlated,worm2022}. For more details about the calculation settings,  see Appendix~\ref{app:calc}.

For consistency and to be able to draw direct comparisons, all the materials were studied within a pseudo-tetragonal crystal setting at ambient pressure, with a \textit{P4/mmm} structure being used for La$_{3}$Ni$_{2}$O$_{7}$-1313, and an \textit{I4/mmm} structure for both La$_2$NiO$_4$ and La$_{4}$Ni$_{3}$O$_{10}$. At present, the crystal structure of La$_{3}$Ni$_{2}$O$_{7}$-1313 remains ambiguous, with two potential space groups being more likely at ambient pressure: \textit{Cmmm} (corresponding to a $\sqrt{2}\times\sqrt{2}$ supercell of the tetragonal--$P4/mmm$ parent phase, containing no octahedral tilts)~\cite{chen2023polymorphism, Wang2024longrange}, and \textit{Fmmm}~\cite{puphal2023unconventional} (both exhibiting octahedral tilts).
The crystal structure of La$_{2}$NiO$_{4}$ transitions from a high-temperature tetragonal (\textit{I4/mmm}) phase to a low-temperature orthorhombic (\textit{Bmab}) phase~\cite{Rodriguez-Carvajal1991, Skinner2003} with only subtle differences in the basic correlated features of their electronic structures~\cite{lechermann2022assessing}. For La$_{4}$Ni$_{3}$O$_{10}$, the crystal structure has been resolved as monoclinic (\textit{P2$_1$/a}) or orthorhombic (\textit{Bmab}) at ambient pressure \cite{Zhang2020oxygen}. However, previous works~\cite{jung2022RPs, labollita2024electronic} have shown that the near Fermi level DFT dispersions of the monoclinic phase can be well reproduced by a $\sqrt{2}$$\times$$\sqrt{2}$ tetragonal (\textit{I4/mmm}) structure.
The application of pressure tends to `tetragonalize' the crystal structure of all of the materials studied here, as shown before for the $n=2$ and 3 RP phases \cite{labollita2023electronic,labollita2024electronic,puphal2023unconventional, wang2023i4mmmexp, geisler2023structure}. Hence, the ambiguity in terms of the space group symmetry is removed in the pressurized phases.   

\begin{figure*}
    \centering
    \includegraphics[width=2\columnwidth]{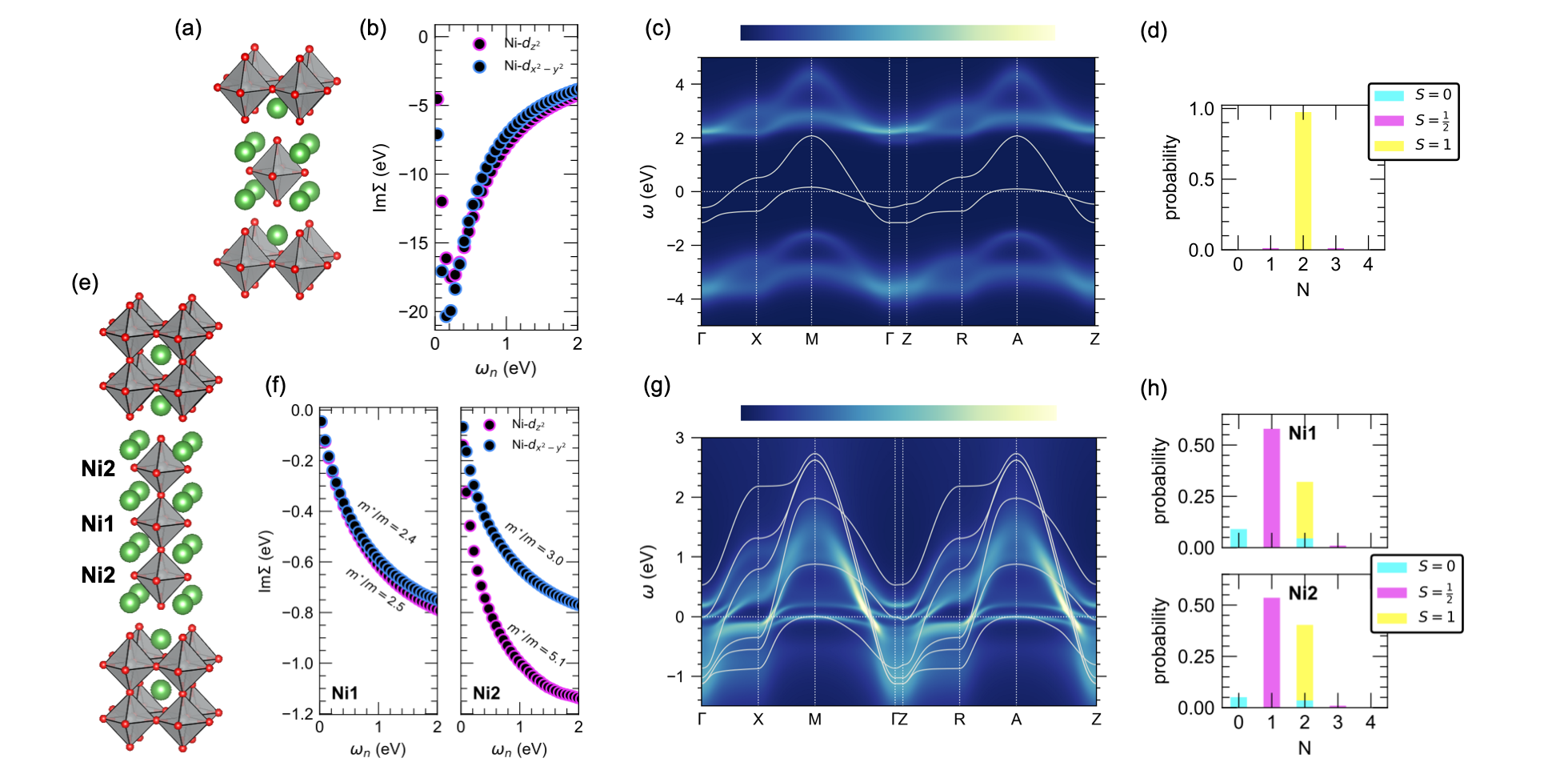}
    \caption{Crystal structure and correlated low-energy electronic structure of La$_{2}$NiO$_{4}$ (a-d) and La$_{4}$Ni$_3$O$_{10}$ (e-h) both in a tetragonal $I4/mmm$ structure at ambient pressure ($T = 116$ K). (a) Crystal structure of La$_{2}$NiO$_{4}$ where green, grey, and red spheres denote La, Ni, and O atoms, respectively.  (b) Imaginary part of the Ni-$e_{g}$ self-energies on the Matsubara axis. (c) $\mathbf{k}$-resolved spectral function $A(\mathbf{k},\omega$) (eV$^{-1}$). (d) Atomic histogram for the Ni-$e_{g}$ impurity from the impurity density matrix (measured within the CT-HYB impurity solver) resolved into particle number ($N$) and spin ($S$). (e) Crystal structure of  La$_{4}$Ni$_{3}$O$_{10}$ with the two inequivalent Ni sites: inner (Ni1) and outer (Ni2) denoted. (f) Imaginary part of the Ni(1,2)-$e_{g}$ self-energies on the Matsubara axis (orbital-resolved mass enhancements computed from the low-frequency part inset). (g) $\mathbf{k}$-resolved spectral function $A(\mathbf{k},\omega$) (eV$^{-1}$). (h) Site-resolved atomic histograms for the Ni-$e_{g}$ impurities resolved into particle number ($N$) and spin ($S$). Solid white lines in (c,g) correspond to the non-interacting bands from DFT.} 
    \label{fig:constituents-P0}
\end{figure*}

\begin{figure*}
\centering
\includegraphics[width=2\columnwidth]{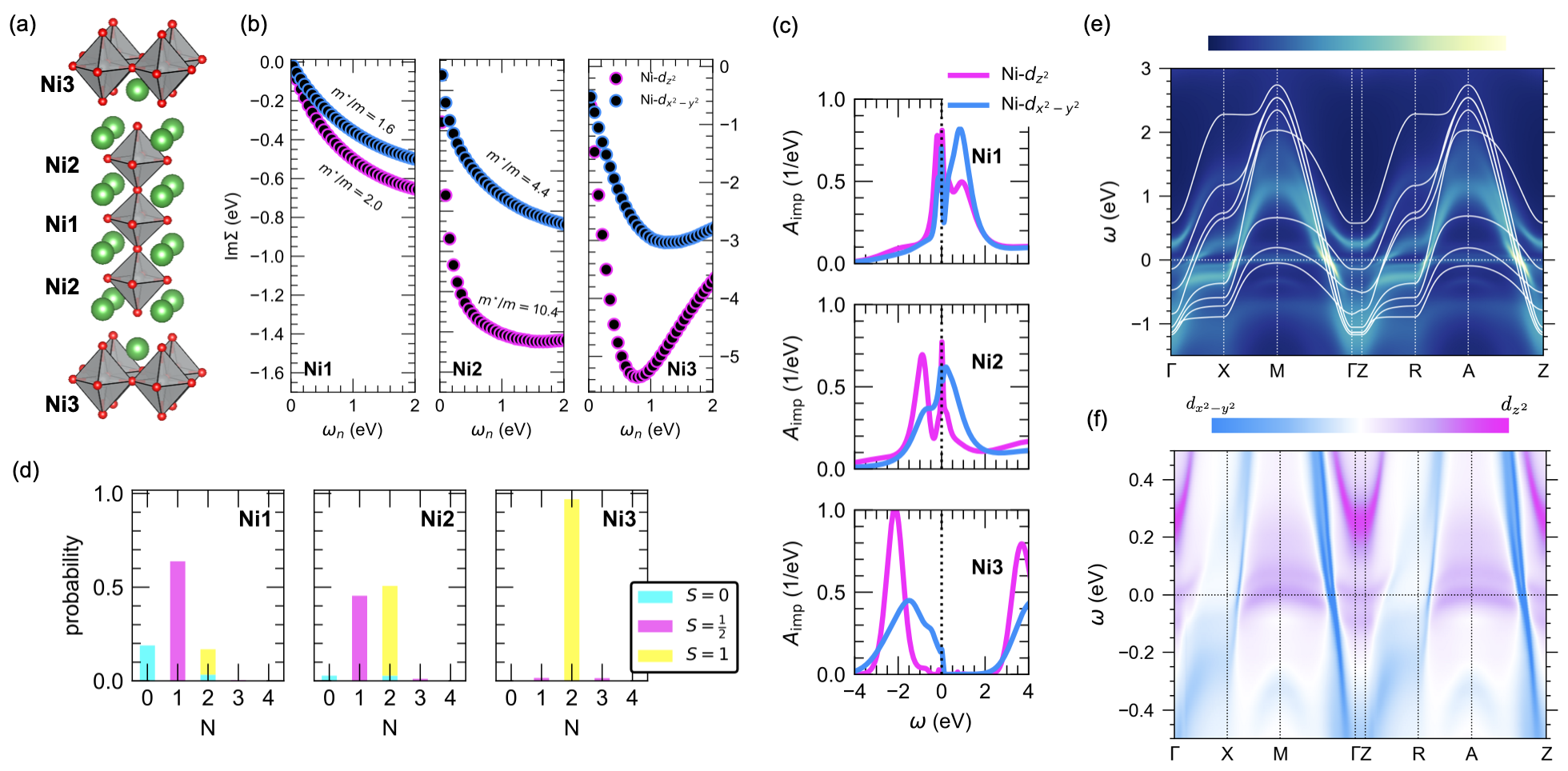}
\caption{Crystal structure and correlated low-energy electronic structure of La$_{3}$Ni$_{2}$O$_{7}$-1313 in a tetragonal \textit{P4/mmm} structure at ambient pressure ($T = 116$ K). (a) Crystal structure of La$_{3}$Ni$_{2}$O$_{7}$-1313  where green, grey, and red spheres denote the La, Ni, and O atoms, respectively, with the three inequivalent Ni sites: inner (Ni1), outer (Ni2), and single-layer (Ni3) denoted. (b) Imaginary part of the Ni(1,2,3)-$e_{g}$ self-energies on the Matsubara axis (orbital-resolved mass enhancements computed from the low-frequency part inset). (c) Site- and orbital (Ni-$e_{g}$)-resolved impurity $\mathbf{k}$-summed spectral data. (d) Site-resolved atomic histograms for the Ni-$e_{g}$ impurities obtained from the impurity density matrix (measured within the CT-HYB impurity solver) resolved into particle number ($N$) and spin ($S$). (e) $\mathbf{k}$-resolved spectral function $A(\mathbf{k},\omega$) (eV$^{-1}$). Solid white lines correspond to the non-interacting bands from DFT. (f) Low-energy blow-up of (e) projected onto the Ni-$e_{g}$ orbitals. Blue (pink) corresponds to Ni-$d_{x^{2}-y^{2}}$ (Ni-$d_{z^{2}}$).}
\label{fig:1313-p0-dmft}
\end{figure*}

\section{\label{sec:results}Results}

\textit{Correlated electronic structure of the structural constituents: La$_{2}$NiO$_{4}$ and La$_{4}$Ni$_{3}$O$_{10}$.--} We start by describing the electronic structure of the two structural constituents of La$_{3}$Ni$_{2}$O$_{7}$-1313, namely, single-layer La$_{2}$NiO$_{4}$ and trilayer La$_{4}$Ni$_{3}$O$_{10}$  at ambient pressure.  Figure ~\ref{fig:constituents-P0} summarizes the key electronic structure features for both materials. The single-layer nickelate La$_{2}$NiO$_{4}$ (see Fig.~\ref{fig:constituents-P0}a) with an average $d^{8}$ filling, exhibits metallic behavior at the non-interacting (DFT) level with two bands of Ni-$e_{g}$ character crossing the Fermi energy (thin, white lines in Fig.~\ref{fig:constituents-P0}c).  However, when including local electronic interactions within (single-site) DMFT, La$_{2}$NiO$_{4}$ transfers into a Mott-insulating regime without the need for antiferromagnetic order (see Fig.~\ref{fig:constituents-P0}b), with a gap opening in both orbital sectors evidenced by the local self-energies shown in Fig.~\ref{fig:constituents-P0}b. The large  ($\sim 4$ eV) gap of this Mott insulating state can be clearly observed in the $\mathbf{k}$-summed spectral data (see Fig. \ref{fig:ksum0} of Appendix \ref{app:p0}) as well as in the  $\mathbf{k}$-resolved spectral function (see Fig.~\ref{fig:constituents-P0}c) that is consistent with photoemission data~\cite{la214photoemission}, and with previous theoretical calculations~\cite{lechermann2022assessing}. Moreover, this correlated insulating state is characterized by a high-spin ($S=1$) configuration within the Ni-$e_{g}$ manifold (see Fig.~\ref{fig:constituents-P0}d) obtained from the impurity density matrix (measured within CT-HYB). This spin configuration corresponds to one electron in each orbital, as expected for a $d^8$ Ni in an octahedral environment (see Table~\ref{tab:orbitals} for the corresponding orbital occupations derived from the impurity Green's functions).

\begin{figure}
    \centering
    \includegraphics[width=\columnwidth]{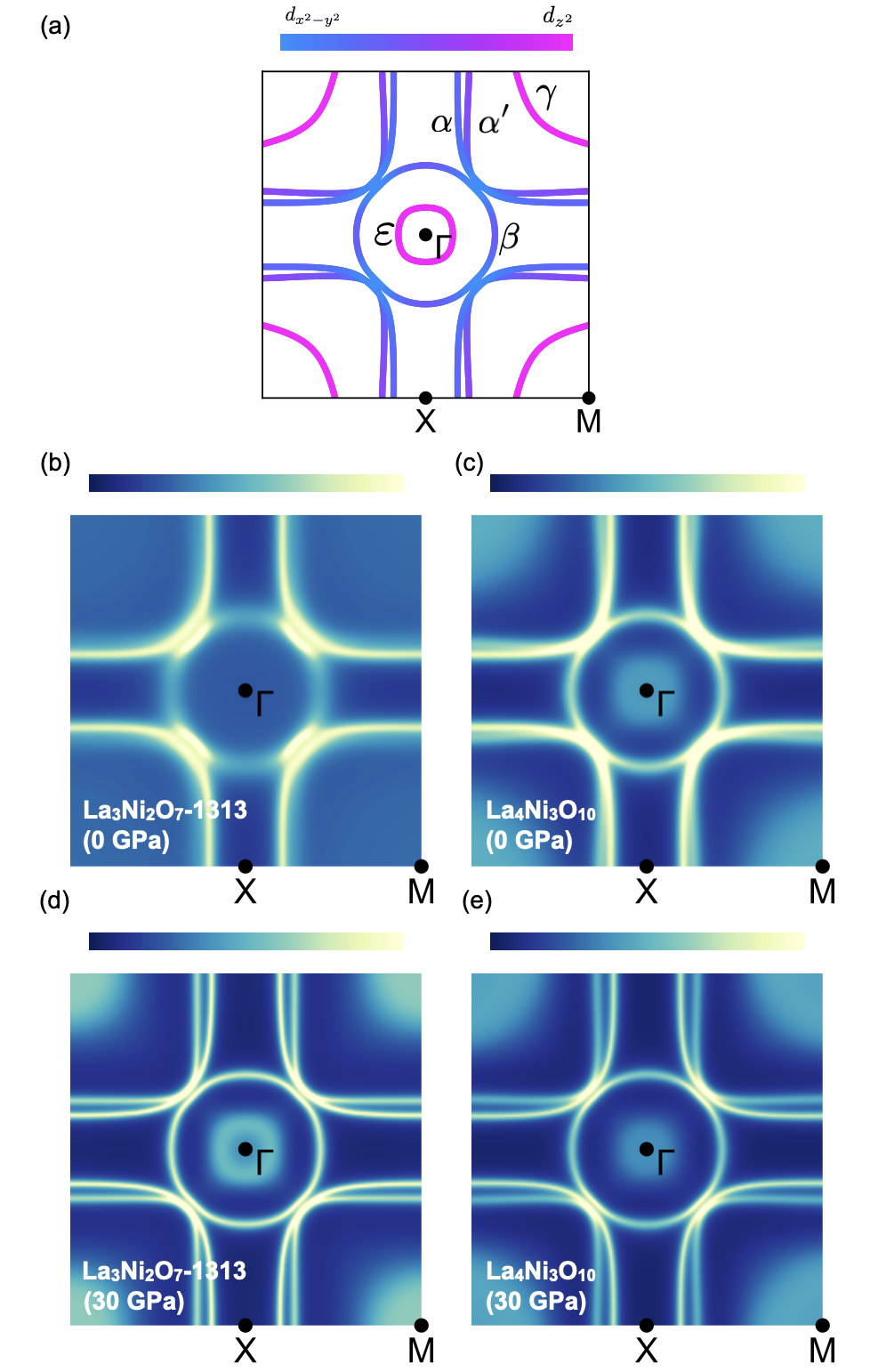}
    \caption{Schematic of the Fermi surface and interacting Fermi surfaces in the $k_{z}=0$ plane for La$_{3}$Ni$_{2}$O$_{7}$-1313 and La$_{4}$Ni$_{3}$O$_{10}$ (a) Schematic fermiology for La$_{3}$Ni$_{2}$O$_{7}$-1313 and La$_{4}$Ni$_{3}$O$_{10}$ highlighting the possible Fermi surface sheets with Ni-$e_{g}$ orbital character denoted. Lowercase Greek letters are used to label the different sheets. The orbital character is highlighted where blue (pink) corresponds to Ni-$d_{x^{2}-y^{2}}$ (Ni-$d_{z^{2}}$) orbital weight. (b,c) Interacting Fermi surfaces  ($T = 116$ K) at ambient pressure and (d,e) at 30 GPa for La$_{3}$Ni$_{2}$O$_{7}$-1313 and La$_{4}$Ni$_{3}$O$_{10}$, respectively.}
    \label{fig:fermisurfaces}
\end{figure}

\begin{table}
    \centering
    \begin{tabular*}{\columnwidth}{c|@{\extracolsep{\fill}}ccc}
    \hline\hline
                     &  $n(d_{z^{2}})$  & $n(d_{x^{2}-y^{2}})$ & $n_{\mathrm{total}}$ \\
    \hline
    La$_{2}$NiO$_{4}$  &                  &                 &               \\
    Ni                 &   1.0 (1.0)  & 1.0 (1.0)           &   2.0 (2.0)  \\
    \hline
    La$_{4}$Ni$_{3}$O$_{10}$  &                  &             &            \\
    Ni1                &   0.64 (0.62)   & 0.61 (0.57)         &   1.25 (1.19) \\
    Ni2                &   0.77 (0.77)   & 0.60 (0.63)          &   1.37 (1.4) \\
    \hline
    La$_{3}$Ni$_{2}$O$_{7}$-1313 &                  &             &            \\
    Ni1                &   0.57 (0.62)   & 0.43 (0.56)         &   1.0 (1.18)  \\
    Ni2                &   0.86 (0.81)  & 0.64 (0.63)        &   1.5 (1.44)  \\
    Ni3                &   1.0 (1.0)  &  1.0 (0.94)         &   2.0 (1.94)  \\
    \hline\hline
    \end{tabular*}
    \caption{Site-resolved Ni-$e_{g}$ orbital occupations for La$_{2}$NiO$_{4}$, La$_{4}$Ni$_{3}$O$_{10}$, and La$_{3}$Ni$_{2}$O$_{7}$-1313 at ambient pressure (30 GPa) obtained from the impurity Green's function.}
    \label{tab:orbitals}
\end{table}

The trilayer RP nickelate La$_{4}$Ni$_{3}$O$_{10}$ (see Fig.~\ref{fig:constituents-P0}e), with an average $d^{7.33}$ filling, can instead be described as a correlated metal with strong layer-dependent electronic correlations when comparing the inner (Ni1) and outer (Ni2) symmetry inequivalent layers, in agreement with Ref. \cite{wang2024nonfermi}. These layer-dependent correlations are revealed by the site-resolved local Ni-$e_{g}$ self-energies shown in Fig.~\ref{fig:constituents-P0}f with the estimated mass enhancements $m^{\star}/m_{\mathrm{DFT}}= (1-\partial\mathrm{Im}\Sigma/\partial\omega_{n}|_{\omega_{n}\rightarrow0})$ from the lowest Matsubara frequencies being notably different. On the inner layers (Ni1), correlations are nearly identical in both orbital sectors with $m^{\star}/m_{\mathrm{DFT}} \sim 2.5$. The outer layers (Ni2) show overall significantly stronger electronic correlations with $m^{\star}/m_{\mathrm{DFT}} \sim 5$ for $d_{z^{2}}$ and $m^{\star}/m_{\mathrm{DFT}} \sim 3$ for the $d_{x^{2}-y^{2}}$ orbitals. Averaging mass enhancements (over layers and orbitals) gives a value of $\sim 3$, which is in agreement with estimated mass enhancements from angle-resolved photoemission spectroscopy (ARPES) data for La$_{4}$Ni$_{3}$O$_{10}$~\cite{Li2017Fermiology}. The layer differentiation  described above can be connected to the differences in orbital occupations between the two layers derived from the impurity Green's functions (see Table~\ref{tab:orbitals}). The filling of the Ni-$d_{x^{2}-y^{2}}$ orbitals is nearly identical on the two layers, while for the $d_{z^2}$ orbitals Ni2 (outer) is closer to half-filling (with $n_{d_{z^{2}}}=0.77$) than Ni1 (inner), with $n_{d_{z^{2}}}=0.64$.  The total layer-resolved occupations for Ni1 (inner), Ni2 (outer) read: $n = 1.25, 1.37$, respectively. These differences in occupations result in differences in the preferred atomic configurations. The inner layer slightly disfavors a high-spin state relative to the outer layer (see Fig.~\ref{fig:constituents-P0}h). Albeit subtle in our calculations, this tendency to penalize a high-spin state in the inner layer is in line with the experimentally observed spin density-wave state~\cite{Zhang2020Intertwined}, consistent with non-magnetic inner layers and magnetic outer layers (such a state has also been described theoretically within DFT+$U$~\cite{labollita2024electronic} and RPA-based calculations~\cite{zhang2024prediction}). The $\mathbf{k}$-resolved spectral function in Fig.~\ref{fig:constituents-P0}g together with the $\mathbf{k}$-summed spectral data in Fig. \ref{fig:ksum0} of Appendix \ref{app:p0} show that the low-energy physics is dominated by the Ni-$e_{g}$ states with strong orbital mixing. The non-interacting DFT dispersions (thin, white lines in Fig.~\ref{fig:constituents-P0}g) appear strongly renormalized due to correlations. Specifically, flat band features of mostly Ni-$d_{z^{2}}$ character from both layers flank the chemical potential from above and below with the remaining near chemical potential spectral weight exhibiting Ni-$d_{x^{2}-y^{2}}$ character. The main features that we just described for the correlated electronic structure of La$_{4}$Ni$_{3}$O$_{10}$ in a tetragonal \textit{I4/mmm} phase remain in the monoclinic \textit{P2$_1$/a} structure as shown in Fig. \ref{fig:4310-p21a-summary} of Appendix \ref{app:p21a}.

\begin{figure*}
\centering
\includegraphics[width=2\columnwidth]{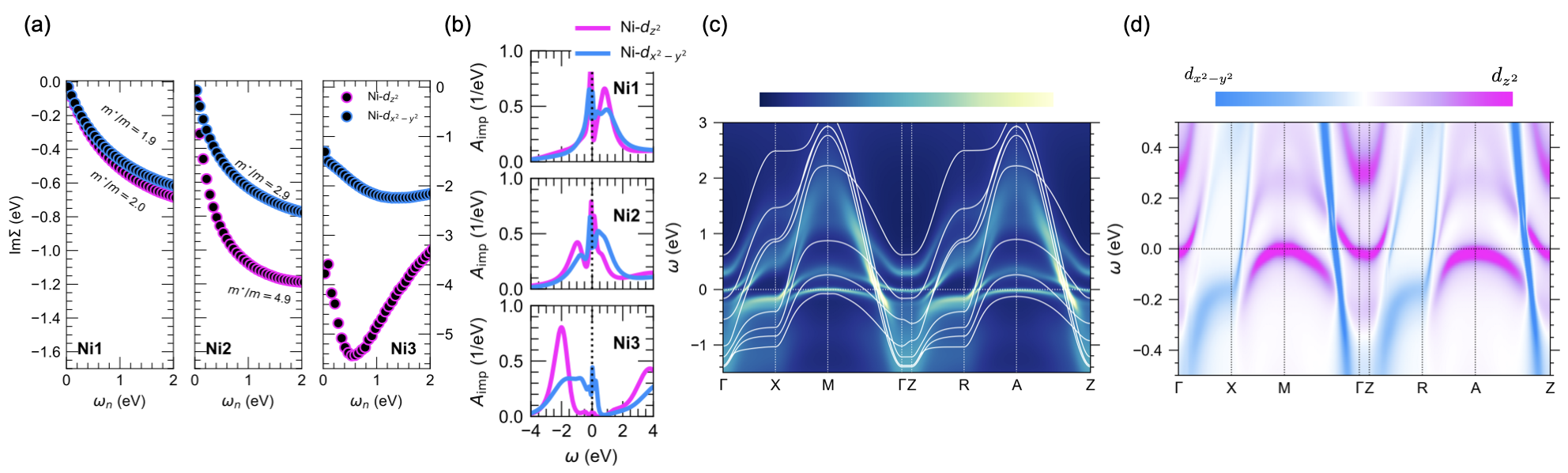}
\caption{Correlated low-energy electronic structure of La$_{3}$Ni$_{2}$O$_{7}$-1313 at 30 GPa ($T = 116$ K) for the corresponding DFT-relaxed structure (tetragonal, \textit{P4/mmm}). 
(a) Imaginary part of the Ni(1,2,3)-$e_{g}$ self-energies on the Matsubara axis (orbital-resolved mass enhancements computed from the low-frequency part inset). (b) Site- and orbital (Ni-$e_{g}$)-resolved impurity $\mathbf{k}$-summed spectral data. (c) $\mathbf{k}$-resolved spectral function $A(\mathbf{k},\omega$) (eV$^{-1}$). Solid white lines correspond to the non-interacting (DFT) band structure. (d) Low-energy blow-up of (c) projected onto the Ni-$e_{g}$ orbitals. Blue (pink) corresponds to Ni-$d_{x^{2}-y^{2}}$ (Ni-$d_{z^{2}}$).}
\label{fig:1313-p30-dmft}
\end{figure*}

\textit{Correlated electronic structure of the La$_{3}$Ni$_{2}$O$_{7}$-1313 polymorph.--} We now turn to the low-energy correlated electronic structure of La$_{3}$Ni$_{2}$O$_{7}$-1313 (with an average $d^{7.5}$ filling) at ambient pressure, which is summarized in Fig.~\ref{fig:1313-p0-dmft}. As explained above, the structure of La$_{3}$Ni$_{2}$O$_{7}$-1313 is a structural superposition of the single-layer and trilayer nickelates stacked along the $c$-axis (see Fig.~\ref{fig:1313-p0-dmft}a)~\cite{chen2023polymorphism, puphal2023unconventional}. This structural amalgam renders three inequivalent Ni sites: Ni1 and Ni2 in the inner and outer layers of the trilayer, respectively, and Ni3 in the single-layer block.  Overall, the main features of the correlated electronic structure for La$_{3}$Ni$_{2}$O$_{7}$-1313 can be easily traced to its underlying structural counterparts described in the previous section. 
Starting with the single-layer, we find that it is in a Mott-insulating regime (with gaps in both orbital sectors), as shown in Figs.~\ref{fig:1313-p0-dmft}b,c,  with one electron in each $e_g$ orbital (see Table~\ref{tab:orbitals}). This gives rise to a high-spin ($S=1$) configuration as shown in Fig.~\ref{fig:1313-p0-dmft}d. These findings are in agreement with other DFT+DMFT calculations ~\cite{lechermann2024electronic}.  As the single-layer is in this Mott insulating regime, the low-energy physics of La$_{3}$Ni$_{2}$O$_{7}$-1313 is subsequently dominated by the trilayer block (see Fig.~\ref{fig:1313-p0-dmft}c). Within the trilayer, both layers can be described as correlated metals with correlations being stronger on the outer layers (Ni2) than the inner layers (Ni1), identical to the nominal trilayer La$_{4}$Ni$_{3}$O$_{10}$ RP described above (see Fig.~\ref{fig:1313-p0-dmft}b). While the layer-dependence is qualitatively similar to the trilayer, the local Ni-$e_{g}$ self-energies shown in Fig.~\ref{fig:1313-p0-dmft}b indicate that some quantitative differences exist. For the outer Ni sites (Ni2), the Ni-$d_{z^{2}}$ orbital is much more correlated in La$_{3}$Ni$_{2}$O$_{7}$-1313 with a mass renormalization $m^{\star}/m_{\mathrm{DFT}}\sim 10$ indicating that these states are more localized in La$_{3}$Ni$_{2}$O$_{7}$-1313 than in La$_{4}$Ni$_{3}$O$_{10}$. The intensified correlations on the outer layers of the trilayer can be understood from its proximity to a correlation-driven insulating layer, forcing the Ni2-$d_{z^{2}}$ states to localize -- this feature is absent in La$_{4}$Ni$_{3}$O$_{10}$. The similar layer-dependence between the individual constituents and La$_{3}$Ni$_{2}$O$_{7}$-1313 is also obvious when comparing the $\mathbf{k}$-summed spectral data in Fig. \ref{fig:ksum0} of Appendix \ref{app:p0}. The layer-resolved orbital occupation data in La$_{3}$Ni$_{2}$O$_{7}$-1313 reads: $n=1$ (Ni1), $n=1.5$ (Ni2), and $n=2$ (Ni3) (see Table~\ref{tab:orbitals}). These occupations are expected given the average Ni$^{2.5+}$ valence for La$_{3}$Ni$_{2}$O$_{7}$-1313 and further highlight the large degree of layer differentiation in this material. The $\mathbf{k}$-resolved spectral function is shown in Figs.~\ref{fig:1313-p0-dmft}e,f. At the non-interacting level, several bands cross the Fermi energy with contributions from all layers and both Ni-$e_{g}$ orbitals (thin, white lines in Fig.~\ref{fig:1313-p0-dmft}e). We find that local interactions significantly renormalize the bare (DFT) band dispersions and redistribute spectral weight away from the chemical potential (particularly, spectral weight from the single-layer block as a consequence of the opening of a Mott gap). A low-energy blow up of the spectral data projected onto the Ni-$e_{g}$ orbitals in Fig.~\ref{fig:1313-p0-dmft}f reveals strongly decoherent Ni-$d_{z^{2}}$ states around the chemical potential. Overall, our results for La$_{3}$Ni$_{2}$O$_{7}$-1313 in a tetragonal (\textit{P4/mmm}) setting are in good agreement with many-body electronic structure calculations performed in the \textit{Fmmm} crystal structure with octahedral tilts \cite{lechermann2024electronic}.

Good agreement can be observed when comparing our $\mathbf{k}$-dependent spectral data to the available ARPES experiments for La$_{3}$Ni$_{2}$O$_{7}$-1313~\cite{abadi2024electronic} and La$_4$Ni$_3$O$_{10}$ \cite{Li2017Fermiology}, especially when examining the corresponding interacting Fermi surfaces (see Fig.~\ref{fig:fermisurfaces}). At ambient pressure, the fermiology of both materials is characterized by large hole-like pockets of dominant Ni-$d_{x^{2}-y^{2}}$ character ($\alpha$, $\alpha'$ sheets in the schematic of Fig.~\ref{fig:fermisurfaces}a) centered at the corner (M) of the Brillouin zone, split due to the interlayer coupling within the trilayer (see Figs.~\ref{fig:fermisurfaces}b,c). For La$_{4}$Ni$_{3}$O$_{10}$, a coherent circular electron-like pocket ($\beta$) with mixed orbital character is also present, while incoherent spectral weight appears in La$_{3}$Ni$_{2}$O$_{7}$-1313 in this part of the zone. Incoherent spectral weight of Ni-$d_{z^{2}}$ character is also present in La$_{4}$Ni$_{3}$O$_{10}$ at the zone corner and zone center (in the position of the $\gamma$ and $\varepsilon$ pockets in the schematic) due to the strong electronic correlations in this orbital sector. This spectral weight is absent in La$_{3}$Ni$_{2}$O$_{7}$-1313 at ambient pressure. The nearby flat band of $d_{z^2}$ character observed by ARPES \cite{abadi2024electronic} below (and close) to the Fermi energy in La$_{3}$Ni$_{2}$O$_{7}$-1313 can also be observed in our spectral data (see Fig.~\ref{fig:1313-p0-dmft}e). 

We now analyze the evolution of the low-energy correlated electronic structure under pressure. Signatures of superconductivity in La$_{3}$Ni$_{2}$O$_{7}$-1313 emerge above 10 GPa and in La$_4$Ni$_3$O$_{10}$ above 20-30 GPa. In order to perform direct comparisons between these two materials, we analyze their electronic structure under pressure using DFT-relaxed structures at 30 GPa which is well within the superconducting regime for both compounds. We focus in the main text on the results for La$_{3}$Ni$_{2}$O$_{7}$-1313 while further details about the correlated electronic structure of the structural constituents at 30 GPa can be found in Figs.~\ref{fig:constituents-p30} and ~\ref{fig:ksum30} of Appendix~\ref{app:p30}. For pressurized La$_{3}$Ni$_{2}$O$_{7}$-1313, we find that the Mott-insulating nature of the single-layer block at ambient pressure transitions into an orbital-selective Mott state with the Ni-$d_{x^{2}-y^{2}}$ orbital exhibiting metallic behavior (see Figs.~\ref{fig:1313-p30-dmft}a,b). Electronic correlations weaken within the trilayer with the outer layers still being the most correlated (see Fig.~\ref{fig:1313-p30-dmft}a) with $m^{\star}/m_{\mathrm{DFT}} \sim 2$ for both orbitals on Ni1 (inner) and $m^{\star}/m_{\mathrm{DFT}} \sim 3,5$ for the Ni-$d_{x^{2}-y^{2}}$, $d_{z^{2}}$ orbitals, respectively, on Ni2 (outer). Both changes in the layer-resolved physics (with respect to ambient pressure) can be understood from the increased electronic bandwidth with pressure, which changes the critical ratio $U/t$. Interestingly, pressure induces a charge redistribution between the single-layer and trilayer units with charge flowing from the single-layer (Ni3) into the trilayer (Ni1,Ni2).  The total occupations on the three distinct layers read: 1.18, 1.44, and 1.94 for Ni1, Ni2, and Ni3, respectively (see Table~\ref{tab:orbitals}). Atomic histograms (not shown) remain qualitatively similar to the ambient pressure case (see Fig.~\ref{fig:1313-p0-dmft}c), with only subtle reshuffling of probabilities between different configurations. Overall, the application of pressure effectively electron dopes the trilayer block above the nominal filling of 4 $e_g$ electrons per trilayer to 4.06 electrons per trilayer, about a 2\% doping. A similar charge-transfer, self-doping effect is also described in Ref.~\cite{zhang2024electronic} using alternative theoretical methods. This pressure-induced electron doping may have consequences on the superconducting transition temperature of La$_{3}$Ni$_{2}$O$_{7}$-1313 (T$_c$ $\sim$ 80 K) relative to La$_{4}$Ni$_{3}$O$_{10}$ (T$_c$ $\sim$ 30 K). Consistent with this observation, previous theoretical calculations for La$_{4}$Ni$_{3}$O$_{10}$ indeed showed that the superconducting correlations in all leading pairing channels ($s^{\pm}$, $d_{xy}$, $d_{x^{2}-y^{2}}$) increase with electron doping~\cite{zhang2024prediction,yang2024effective}. 

Overall, the $\mathbf{k}$-dependent spectral data shows that the low-energy physics of pressurized La$_{3}$Ni$_{2}$O$_{7}$-1313 continues to be governed by strongly hybridized Ni-$e_{g}$ states with the additional contribution from the metallic single-layer $d_{x^{2}-y^{2}}$ orbital. Compared to the ambient pressure case, the Ni-$d_{z^{2}}$ bands become much more coherent with sharper `flat band-like' spectral features around the chemical potential (see Figs.~\ref{fig:1313-p30-dmft}c,d and Fig. \ref{fig:ksum30} in Appendix \ref{app:p30}). When looking at the Fermi surface of La$_{3}$Ni$_{2}$O$_{7}$-1313 at 30 GPa (see Fig.~\ref{fig:fermisurfaces}d) some of the features described above at ambient pressure remain but appear more coherent, in particular the two hole-like pockets of dominant Ni-$d_{x^{2}-y^{2}}$ character around M ($\alpha$, $\alpha'$), and the $\beta$ pocket of mixed orbital character at $\Gamma$  (analogous to the pressurized  La$_{4}$Ni$_{3}$O$_{10}$ material as shown in Fig.~\ref{fig:fermisurfaces}e).  As the coherence of the Ni-$d_{z^{2}}$ states also increases, clear spectral weight can be observed at the zone corner ($\gamma$ pocket) and zone center ($\varepsilon$ pocket) for both materials.  Because of the charge-transfer from the single-layer to the trilayer in La$_{3}$Ni$_{2}$O$_{7}$-1313, the effective electron filling within the trilayer is larger than in La$_{4}$Ni$_{3}$O$_{10}$, which increases the size of the electron-pocket at the zone center and decreases the size of the hole-pockets at the zone corners relative to nominal La$_{4}$Ni$_{3}$O$_{10}$. Overall, with the above considerations, the interacting fermiology of pressurized La$_{3}$Ni$_{2}$O$_{7}$-1313 can simply be interpreted as electron-doped La$_{4}$Ni$_{3}$O$_{10}$~\cite{geisler2024opticalpropertieselectroniccorrelations}.

\section{\label{sec:summary}Summary and discussion} 
Our DFT+DMFT calculations for La$_{3}$Ni$_{2}$O$_{7}$-1313 (and its structural constituents La$_{2}$Ni$_{}$O$_{4}$ and La$_{4}$Ni$_{3}$O$_{10}$) elucidate several key features concerning the evolution of the correlated electronic structure of this material under pressure. At ambient pressure, the correlated electronic structure of La$_{3}$Ni$_{2}$O$_{7}$-1313 is characterized by distinctly different physics on the three inequivalent layers with a correlated metallic trilayer unit (characterized by variable correlations on the inner and outer layers) sandwiched by a Mott-insulating single-layer. This alternating stack of insulating and metallic layers renders the correlated physics much stronger in the trilayer block of  La$_{3}$Ni$_{2}$O$_{7}$-1313 than in the nominal trilayer block in La$_{4}$Ni$_{3}$O$_{10}$. Pressurizing La$_{3}$Ni$_{2}$O$_{7}$-1313 has interesting consequences on the correlation physics of this material. Orbital-selective physics emerges within the single-layer with the Ni-$d_{z^{2}}$ orbital remaining in a Mott regime, while the Ni-$d_{x^{2}-y^{2}}$ orbital becomes metallic. Furthermore, pressure induces a charge flow from the single-layer to the trilayer, which effectively electron dopes the trilayer unit. In this manner, pressurized La$_{3}$Ni$_{2}$O$_{7}$-1313 can be thought of as electron-doped La$_{4}$Ni$_{3}$O$_{10}$ when analyzed at the same pressure. This effective electron doping could be connected to the larger superconducting T$_{c}$ of La$_{3}$Ni$_{2}$O$_{7}$-1313 when compared to La$_{4}$Ni$_{3}$O$_{10}$. Overall, we conclude that correlation-driven layer differentiation plays a crucial role in the electronic structure of the La$_{3}$Ni$_{2}$O$_{7}$-1313 polymorph, closely aligning its low-energy physics with that of the trilayer La$_{4}$Ni$_{3}$O$_{10}$. Electron doping  La$_{4}$Ni$_{3}$O$_{10}$ as well as La$_{3}$Ni$_{2}$O$_{7}$-1313 could be a promising route to increase the superconducting T$_c$ of these materials.

\section*{acknowledgements}
The authors acknowledge support from NSF Grant No. DMR-2045826 and NSF Grant No. DMR-2323971, as well as the ASU Research Computing Center for HPC resources. 

\newpage 
\appendix

\section{\label{app:calc}Structural data and calculation settings}

Crystal structure data for La$_{2}$NiO$_{4}$ ($I4/mmm$), La$_{3}$Ni$_{2}$O$_{7}$-1313 ($P4/mmm$), and La$_{4}$Ni$_{3}$O$_{10}$ ($P2_{1}/a$ and $I4/mmm$) at ambient pressure was obtained from Refs.~\onlinecite{Skinner2003, chen2023polymorphism,jung2022RPs}, respectively. The crystal structures under pressure (at 30 GPa) for the three materials were obtained by relaxing the lattice and internal coordinates with an applied external pressure of 30 GPa within the VASP code~\cite{Kresse:1993bz, Kresse:1996kl, Kresse:1999dk} in the nonmagnetic state and with the Perdew-Burke-Ernzerhof (PBE) version~\cite{gga_pbe} of the generalized gradient approximation (GGA) as the exchange-correlation functional. Subsequent density-functional theory calculations within WIEN2k (as described in the main text) were also performed with the Perdew-Burke-Ernzerhof (PBE) version~\cite{gga_pbe} of the generalized gradient approximation as the exchange-correlation functional. The basis set size was set by $RK_{\mathrm{max}}=7$.  Brillouin zone integration was performed on a $21\times21\times21$ $k$-grid for La$_{2}$NiO$_{4}$ ($I4/mmm$) and La$_{4}$Ni$_{3}$O$_{10}$ ($I4/mmm$), a $30\times29\times11$ $k$-grid was used for La$_{4}$Ni$_{3}$O$_{10}$ ($P2_{1}/a$), and a $37\times37\times7$ $k$-grid for La$_{3}$Ni$_{2}$O$_{7}$ ($P4/mmm$).

To construct our quantum impurity problem(s), the Kohn-Sham DFT bands were downfolded onto localized `atomic'-like orbitals with $e_{g}$ \{$d_{z^{2}}$, $d_{x^{2}-y^{2}}$\} symmetry, which are representative of the Ni-$e_{g}$ states, using maximally localized Wannier functions (MLWFs) as implemented in Wannier90~\cite{Arash2014wannier90}, which is interfaced to \textsc{wien}2k via WIEN2WANNIER~\cite{Kunes2010wien2wannier}. The quantum impurity problem(s) (defined by the two-orbital Ni-$e_{g}$ space) were solved within single-site DMFT employing the \texttt{solid\_dmft} package~\cite{soliddmft} built on top of the TRIQS software library~\cite{Parcollet2015triqs}, as explained in the main text. The quantum impurity problem(s) were solved using the CT-HYB impurity solver as implemented in TRIQS/cthyb~\cite{Priyanka2016cthyb} utilizing up to $10^{8}$ cycles to obtain accurate quantum Monte Carlo data. Green's functions were represented in a basis of Legendre polynomials (with 70 coefficients) leading to a smooth self-energy on the Matsubara axis. All calculations were performed at an inverse electronic temperature of $\beta = 100$ eV$^{-1}$ ($T = 116$ K). The ``double-counting'' term was absorbed into the chemical potential, since we worked only in the low-energy space spanned by the Ni-$e_{g}$-like Wannier orbitals. Real-frequency data was obtained by analytic continuation using the maximum-entropy method for the impurity Green's functions and self-energies~\cite{Kraberger2017maxent}. False color plots of the momentum-resolved spectral data share the same scale.

\begin{figure}
    \centering
    \includegraphics[width=\columnwidth]{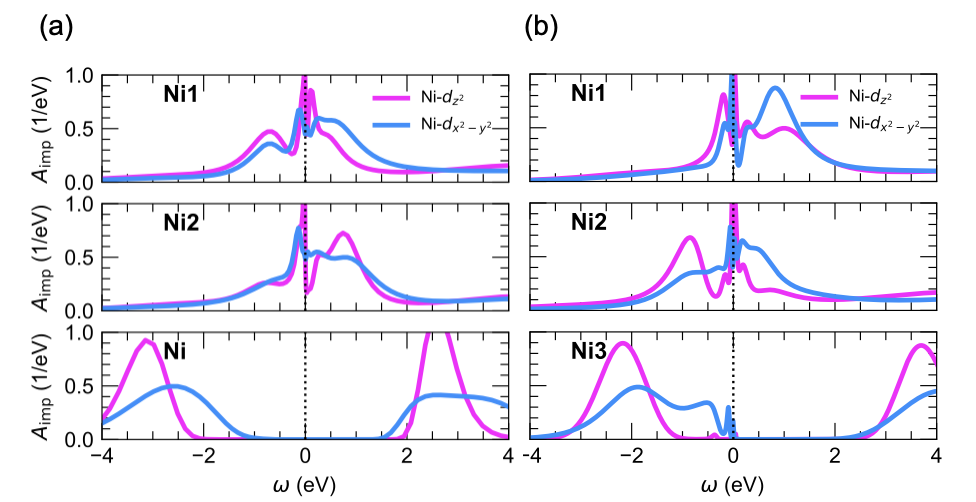}
    \caption{Site- and orbital (Ni-$e_{g}$)-resolved impurity $\mathbf{k}$-summed spectral data at ambient pressure for (a) La$_{2}$NiO$_{4}$ (Ni; bottom panel), La$_{4}$Ni$_{3}$O$_{10}$ (Ni1 and Ni2; top and middle panels), and (b) La$_{3}$Ni$_{2}$O$_{7}$-1313.}
    \label{fig:ksum0}
\end{figure}

\begin{figure*}
    \centering
    \includegraphics[width=2\columnwidth]{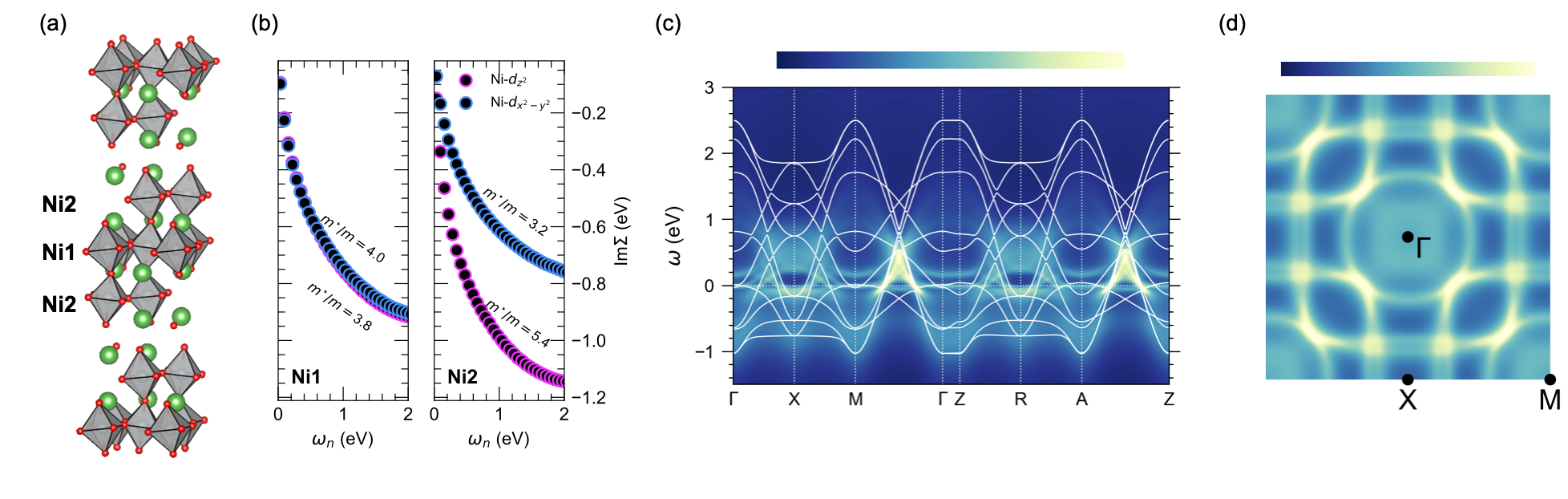}
    \caption{Structure and correlated low-energy electronic structure of La$_{4}$Ni$_{3}$O$_{10}$ in a monoclinic $P2_{1}/a$ setting at ambient pressure ($T = 116$ K). (a) Crystal structure of La$_{4}$Ni$_{3}$O$_{10}$ with the two inequivalent Ni sites: inner (Ni1) and outer (Ni2) denoted. (b) Imaginary part of the Ni(1,2)-$e_{g}$ self-energies on the Matsubara axis (orbital-resolved mass enhancements computed from the low-frequency part inset). (c) $\mathbf{k}$-resolved spectral function $A(\mathbf{k},\omega$) (eV$^{-1}$). Solid white lines correspond to the non-interacting (DFT) band structure. (d) Interacting Fermi surface in the $k_{z}=0$ plane.} 
    \label{fig:4310-p21a-summary}
\end{figure*}

\section{\label{app:p0}Additional data for ambient pressure cases}

Figure~\ref{fig:ksum0} shows the site- and orbital-resolved impurity $\mathbf{k}$-summed spectral data for La$_2$NiO$_4$, La$_4$Ni$_3$O$_{10}$, and La$_3$Ni$_2$O$_7$-1313 
at ambient pressure. The similar layer-dependence between La$_3$Ni$_2$O$_7$-1313 and its structural constituents (with some quantitative differences) is apparent: within the trilayer the low-energy physics is comprised of strongly hybridized Ni-$e_{g}$ states with flat-band like features from the Ni-$d_{z^{2}}$ states and notable differences between inner (Ni1) and outer (Ni2) layers. The single layer (Ni/Ni3) is always in a Mott-insulating state.

\begin{figure*}
   \centering
   \includegraphics[width=2\columnwidth]{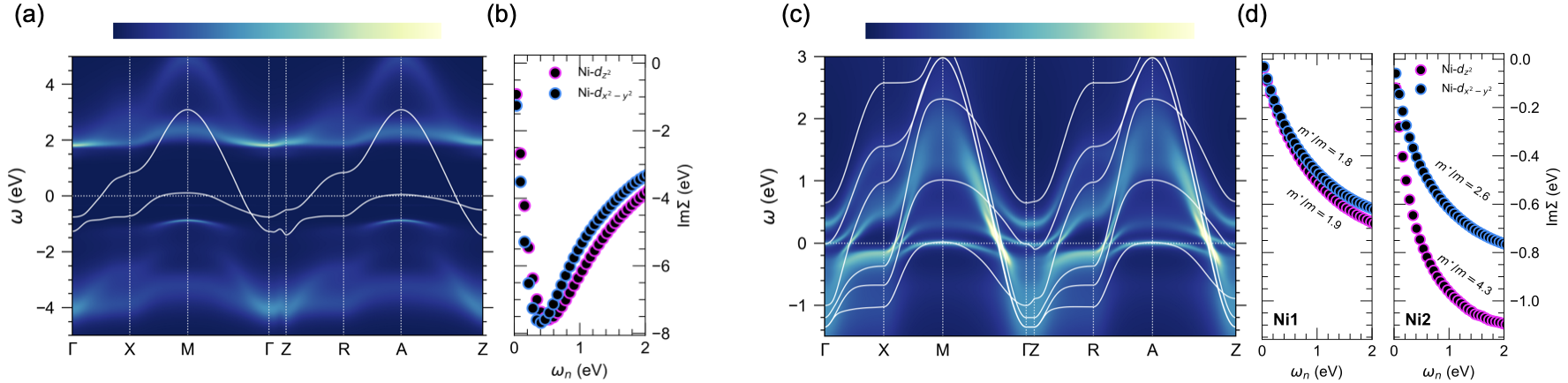}
    \caption{Correlated low-energy electronic structure of La$_2$NiO$_4$ (a-b) and La$_4$Ni$_3$O$_{10}$ (c-d) at 30 GPa ($T = 116$ K) using the corresponding DFT-relaxed structures (both tetragonal, \textit{I4/mmm}).  (a,c) $\mathbf{k}$-resolved spectral function $A(\mathbf{k},\omega$) (eV$^{-1}$) for La$_2$NiO$_4$ (a) and La$_4$Ni$_3$O$_{10}$ (c). Solid white lines correspond to the non-interacting bands from DFT. (b,d) Imaginary part of the Ni-$e_{g}$ self-energies on the Matsubara axis for La$_2$NiO$_4$ (b) and for the two inequivalent Ni atoms in La$_4$Ni$_3$O$_{10}$, Ni1 and Ni2 (d) wherein orbital-resolved mass enhancements are computed from the low-frequency part inset.} 
   \label{fig:constituents-p30}
\end{figure*}

\begin{figure}
    \centering
    \includegraphics[width=\columnwidth]{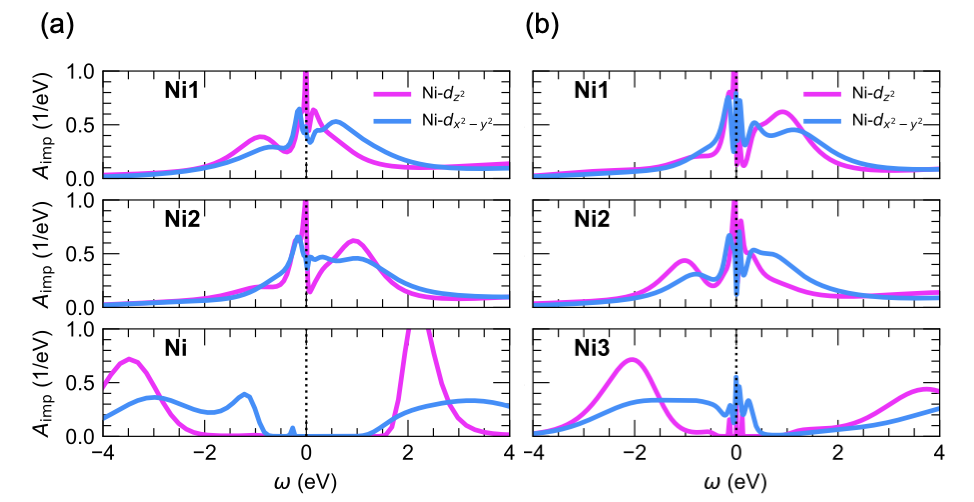}
    \caption{Site- and orbital (Ni-$e_{g}$)-resolved impurity $\mathbf{k}$-summed spectral data at 30 GPa for (a) La$_{2}$NiO$_{4}$ (Ni; bottom panel), La$_{4}$Ni$_{3}$O$_{10}$ (Ni1 and Ni2; top and middle panels), and (b) La$_{3}$Ni$_{2}$O$_{7}$-1313.}
    \label{fig:ksum30}
\end{figure}

\section{\label{app:p21a}Correlated electronic structure of ambient pressure La$_{4}$Ni$_{3}$O$_{10}$ ($P2_{1}/a$)}
Figure~\ref{fig:4310-p21a-summary} summarizes the correlated electronic structure of La$_{4}$Ni$_{3}$O$_{10}$ in the monoclinic ($P2_{1}/a$) crystal setting (see Fig~\ref{fig:4310-p21a-summary}a). Overall, we find qualitative agreement between the main features of the electronic structure of La$_{4}$Ni$_{3}$O$_{10}$ in both monoclinic and tetragonal crystal settings (the latter is described in the main text). The $\mathbf{k}$-resolved spectral data reveals the low-energy physics is comprised of strongly hybridized Ni-$e_{g}$ states with flat-band like features from the Ni-$d_{z^{2}}$ states (see Fig.~\ref{fig:4310-p21a-summary}c). Layer-dependent correlations remain a central correlated feature of La$_{4}$Ni$_{3}$O$_{10}$ even in the monoclinic setting evidenced by the local Ni-$e_{g}$ self-energies (see Fig.~\ref{fig:4310-p21a-summary}b). The topology of the Fermi surface shows `backfolding' compared to the tetragonal crystal structure, however, the main spectral features remain the same (see Fig.~\ref{fig:4310-p21a-summary}d). The atomic multiplets (not shown) for the impurity problem(s) are qualitatively similar to the tetragonal case (see Fig.~\ref{fig:constituents-P0}h in the main text).

\section{\label{app:p30}Additional data for pressurized (P = 30 GPa) cases}
Figures~\ref{fig:constituents-p30} and \ref{fig:ksum30} show the spectral data and local self-energies for La$_{2}$NiO$_{4}$ and La$_{4}$Ni$_{3}$O$_{10}$ at 30 GPa (the structure of both compounds becomes tetragonal under pressure). At 30 GPa, La$_{2}$NiO$_{4}$ remains in a Mott-insulating state with a $\sim 3$ eV charge gap (see Fig.~\ref{fig:constituents-p30}a, and Fig. \ref{fig:ksum30}a), however, the ``Mottness'' is weakened (see Fig.~\ref{fig:constituents-p30}b) compared to the ambient pressure case due to the increased electronic bandwidth of the Ni-$e_{g}$ bands. This is different than the pressurized La$_{3}$Ni$_{2}$O$_{7}$-1313 where the single layer exhibits orbital-selective physics as described in the main text (see also Fig. \ref{fig:ksum30}b).

The trilayer RP La$_{4}$Ni$_{3}$O$_{10}$ at 30 GPa remains a correlated metal with layer-dependent correlations (see Fig.~\ref{fig:constituents-p30}c and Fig. \ref{fig:ksum30}a). Pressure increases the bandwidth of the Ni-$e_{g}$ bands, which decreases the electronic correlations on both layers (see Fig.~\ref{fig:constituents-p30}d). The inner (Ni1) layers always exhibit weaker correlations, than the outer (Ni2) layers in connection to the different orbital fillings between the two layers (see Table~\ref{tab:orbitals}).

\bibliography{refs.bib}

\end{document}